\documentclass[preprint,12pt]{elsarticle}
\usepackage{amssymb}
\usepackage{amsmath}

\journal{Nuclear Instruments and Methods in Physics Research A}

\begin{document}

\begin{frontmatter}

\title{Numerical Simulations of Charge Trapping in Germanium Strip Detectors}

\author[a,b]{Steven E. Boggs\corref{cor1}}
\ead{seboggs@ucsd.edu}
\cortext[cor1]{Corresponding author.}

\author[a]{Sean N. Pike}

\author[a]{Jarred Roberts}
\author[c]{Albert Y. Shih}
\author[b]{John A. Tomsick}
\author[b]{Andreas Zoglauer}

\affiliation[a]{organization={Department of Astronomy \& Astrophysics, University of California, San Diego},
            addressline={9500 Gilman Drive}, 
            city={La Jolla},
            state={CA},
            postcode={92093}, 
            country={USA}}
            
\affiliation[b]{organization={Space Sciences Laboratory, University of California, Berkeley},
            addressline={7 Gauss Way}, 
            city={Berkeley},
            state={CA},
            postcode={94720}, 
            country={USA}}
            
\affiliation[c]{organization={NASA Goddard Space Flight Center},
city={Greenbelt}, state={MD}, postcode={20771}, country={USA}}

\begin{abstract}

Charge trapping in germanium detectors will inevitably impact their excellent spectral performance. Disordered regions in the germanium crystal structure, either created in the material during processing or induced by radiation exposure, will affect the Charge Collection Efficiency (CCE), degrading the spectral resolution. Here we present numerical simulations of charge trapping effects on the anode and cathode signals for cross-strip germanium detectors. We discuss the assumptions behind our model of trapping, which accounts for both the drift length and thermal motion of the charge carriers. We present simulated CCE curves as a function of interaction depth within the detectors, and develop a technique for benchmarking these simulations against measured data.  Comparison with measured CCE curves are presented. We are developing these numerical models with a goal of characterizing, and ultimately correcting, the effects of radiation damage on the spectral resolution of germanium cross-strip detectors. 

\end{abstract}



\end{frontmatter}


\section{Introduction}
\label{sect:intro}

Charge trapping effects in germanium semiconductor detectors (GeDs) have primarily been documented as a result of radiation induced displacement damage (``radiation damage'') to the detectors. Radiation damage effects in GeDs have been extensively studied for fast neutron induced damage due to predominantly neutron exposure in nuclear accelerator experiments \cite{kraner1968effects}. Proton induced damage, which dominates in the space environment, has been studied to a much lesser extent \cite{pehl1978high,koenen1995radiation}. However, both types of exposure produce disordered regions in the crystal structure that act as charge traps, leading to incomplete charge collection and resulting in degraded spectral performance. Radiation damage predominantly results in increased hole trapping in GeDs, with little effect on the electron charge collection \cite{kraner1968effects}. Trapping in undamaged detectors is often assumed to be negligible; however, this is not always true for germanium detectors. Electron trapping appears to dominate over hole trapping for undamaged detectors \cite{hull2014charge,martini1970trapping,mei2020impact}, but both carriers can experience intrinsic trapping to the point of significantly affecting the spectral performance \cite{pike2023}.

\begin{figure}
\centering
\includegraphics[width=0.6\textwidth]{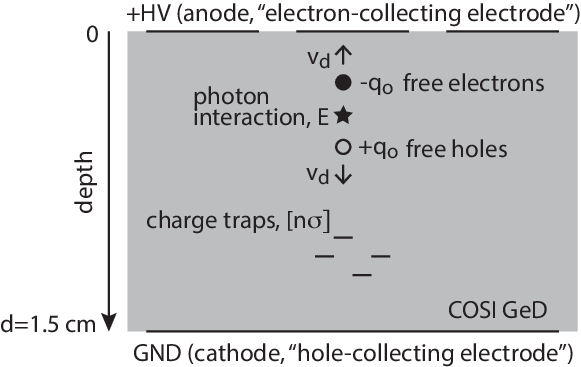}
\caption{\label{fig:f1} Diagram of the charge collection process in COSI GeDs for a photon interaction of energy $E$. The number of free electron-hole pairs is directly proportional to the energy deposited (1 e-h pair per 2.98\,eV). In an applied electric field these e-h pairs will separate and drift in opposite directions, electrons towards the anode and holes towards the cathode. Trapping of these charges as they traverse the detector will result in incomplete collection of these charges on the electrode (CCE $< 1$), leading to an underestimation of the interaction energy unless corrected.}
\end{figure}

The Compton Spectrometer and Imager (COSI) is a soft $\gamma$--ray survey telescope (0.2-5\,MeV) designed to probe the origins of Galactic positrons, reveal sites of ongoing element formation in the Galaxy, use $\gamma$--ray polarimetry to gain insight into extreme environments, and explore the physics of multi-messenger events \cite{kierans20172016,sleator2019benchmarking,arxiv.2109.10403}. The COSI detectors are custom, large-volume (54\,cm$^2$ area, 1.5\,cm thick) cross-strip germanium detectors (Fig.~\ref{fig:f1}) utilizing amorphous contact technologies \cite{amman2007amorphous}. Cross-strip electrodes on the opposite faces, combined with signal timing, provide full 3D position resolution for interactions within the detector. In this work we are focused on modeling intrinsic charge trapping in our original 2.0-mm strip pitch germanium detectors that flew on the COSI balloon payload \cite{kierans20172016,sleator2019benchmarking}. Our ultimate goal is to be able to characterize and correct both the intrinsic and radiation induced charge trapping in the 1.162-mm strip pitch COSI-SMEX detectors as they are exposed to radiation while in orbit over the course of the COSI mission. 

Our goal in this work is to develop numerical simulations of the effects of charge carrier trapping on the Charge Collection Efficiency (CCE) as a function of interaction depth in the COSI cross-strip GeDs, as well as develop a method for benchmarking these simulations against real calibration data. Here we define the CCE as the ratio of the charge ultimately induced on a signal electrode (anode or cathode) relative to the total carrier charge created by the initial photon interaction (drift electrons and holes, respectively). 

In Section~\ref{sect:model}, we describe the physical assumptions behind various methods of modeling charge trapping and discuss how these assumptions apply for germanium detectors. In Section~\ref{sect:test}, we describe implementation of this trapping into our existing charge transport code and verify the precision of our numerical model in regard to modeling the impacts of trapping on the CCE. Section~\ref{sect:CCE} presents simulated results of the effects of trapping as a function of interaction depth in our detectors. In Section~\ref{sect:calibration}, we discuss some aspects of these simulated results leading to a simplified method of benchmarking the simulations against measured data. Section~\ref{sect:average} shows the effects of averaging initial interaction locations across the strips on the resulting CCE curves.  Section~\ref{sect:data} presents a demonstration of benchmarking the simulations against measured data. Finally, we conclude with a discussion of applications for this work in support of the COSI program.

\section{Trapping Model Assumptions}
\label{sect:model}

Trapping of charge carriers in germanium detectors has historically been characterized by a mean free trapping length, $\lambda$ \cite{raudorf1987effect}, which can differ significantly for holes ($\lambda_h$) and electrons ($\lambda_e$). In the presence of a finite mean free trapping length, the fraction of charges still free after traveling a distance $z$ is $q=q_0 e^{-z/\lambda}$. For a detector of characteristic drift dimension $d$, the impact of trapping on the spectral resolution ($\Delta E/E$) becomes significant when $d/\lambda \sim \Delta E/E$.

By contrast, charge trapping in other semiconductor detectors (e.g., silicon, CdZnTe) is usually characterized by a trapping time, $\tau$, which again can differ significantly for holes ($\tau_h$) and electrons ($\tau_e$). For a finite drift time $t$, the fraction of charge carriers still free is $q=q_0 e^{-t/\tau}$. It is common to see the relation $\lambda = v_d \tau$, relating the mean free trapping length $\lambda$ to the trapping time $\tau$, where the drift velocity is a function of electric field, $v_d(E)$. This relation ignores the fundamental assumptions behind adopting $\lambda$ or $\tau$, which hinges on the thermal velocities of the charge carriers as explained below. 

At a basic level, charge traps are characterized by a trap density, $n$, and a trapping cross section, $\sigma$. More details on the origins of these charge traps are provided in \cite{pike2023} and references therein. When a charge has traveled a total path length $l$, the fraction of charge carriers still free is $q=q_0 e^{-l [n \sigma]}$. For a drift time $t$, this total path length is composed of two components: the net drift displacement, $v_d t$, and the thermal path length, $v_{th} t$, where $v_{th}$ is the thermal velocity of the given charge carrier. While the drift induces a net displacement, the net displacement for thermal motion is zero, the same as effectively orthogonal to the drift displacement. Given the vector nature of these motions we cannot simply sum the drift speed and thermal speed to determine the total path length traveled by the charge carrier. Instead, when we add these vectors and average over all relative orientations, the resulting average total path length is determined by adding these two components of speed in quadrature:

\begin{equation}
l = (v_d^2+v_{th}^2)^{\frac{1}{2}} t
\label{eqn:path}
\end{equation}

\noindent This result leads to three separate regimes for modeling charge trapping.

1. Trapping length, $v_d >> v_{th}$.  In this case we can assume that $l \sim v_{d} t$, ignoring the thermal velocities. The fraction of charge carriers free after drifting a distance $z$ would be given by $q=q_0 e^{-z [n \sigma]} = q_0 e^{-z/\lambda}$, where:

\begin{equation}
\lambda =  [n \sigma]^{-1}
\label{eqn:lambda}
\end{equation}

\noindent Note that $[n \sigma]$ is usually assumed uniform over the detector volume, hence $\lambda$ is taken as a single trapping length for the detector. This model of trapping only holds when $v_d  >> v_{th}$. This has commonly been the model used in germanium detectors, but we show below that this implicit assumption about $v_d$ does not hold.

2. Trapping time, $v_d << v_{th}$. By contrast, in this case we can assume that $l \sim v_{th} t$, in which case the fraction of charge carriers free after drifting a time $t$ would be given by $q=q_0 e^{-v_{th} t [n \sigma]} = q_0 e^{-t/\tau}$, where:

\begin{equation}
\tau = [v_{th} n \sigma]^{-1}
\label{eqn:tau}
\end{equation}

\noindent Again, $[n \sigma]$ is usually assumed uniform over the detector volume, and $v_{th}$ is constant for a given charge carrier in a detector at constant temperature. Hence, $\tau$ is taken as a single trapping time for the detector. This model of trapping is only valid when $v_d << v_{th}$, which usually holds for silicon and CdZnTe detectors.

3. Full model, $v_d \sim v_{th}$. In the case where neither of the velocities dominate, the path length needs to be modeled properly by Eqn.~\ref{eqn:path}. In this case, neither $\lambda$ nor $\tau$ can properly model the charge trapping. The correct way to parameterize the trapping is through the product $[n \sigma]^{-1}$ directly, which we will refer to as the \emph{trapping product} to distinguish it from the trapping length, $\lambda$, which has the same units of $[distance]$.  

For charge carriers in a semiconductor at temperature $T$, the root mean square thermal velocities are given by the relation:

\begin{equation}
v_{th} =  \sqrt{\frac{3kT}{m^*}}
\label{eqn:vthermal}
\end{equation}

\noindent Where $m^*$ is the conductivity effective mass for the individual charge carrier. In germanium,  $m^* = 0.12 m_e$ for electrons, and $m^* = 0.21 m_e$ for holes. For an operational temperature of $80$\,K, the corresponding thermal velocities are $v_{th} = 1.7\times10^7$\,cm/s for electrons, and $v_{th} = 1.3\times10^7$\,cm/s for holes. For comparison, drift velocities for both carriers saturate in high electric fields (typical for germanium detector operation) at $\sim 10^7$\,cm/s \cite{bruyneel2006characterization}, comparable to the thermal velocities. Hence the assumption behind using a trapping length, $\lambda$, does not hold for germanium detectors, and the full model utilizing the path length including both the effects of drift velocities and thermal velocities should be utilized to more accurately simulate the effects of trapping in these detectors. 

Regarding the general applicability of the methods developed in this paper, we are utilizing germanium materials with exceptionally low impurity concentrations, typically below $10^{10}$\,cm$^{-3}$. As a result, for a detector thickness $d$, we are in the regime where the trapping product $[n \sigma]^{-1} >> d$. We anticipate this relation to hold even in the case of extensive radiation damage. As such, this work may be considered as relevant in the limit of relatively ``weak'' trapping, though the level of trapping investigated in this paper still has a significant effect on the excellent spectral resolution of GeDs. Specifically we note that the results below would not be applicable in the case where $[n \sigma]^{-1} \sim d$, which often occurs for, e.g., CdZnTe detectors.

We have ignored charge carrier detrapping in this work. Previous work has shown that shallow impurity levels (which lead to detrapping) have no measurable effect on the CCE for germanium detectors operating at liquid nitrogen temperatures and at voltages exceeding the full depletion voltage \cite{martini1970trapping,mei2020impact}, both of which hold for our detectors.

\section{Validating the Precision of our Numerical Simulations}
\label{sect:test}

\begin{table}
\centering
\begin{tabular}{ccc}
\hline
& anode signals & cathode signals\\
\hline
$[n \sigma]^{-1}$ [cm] & $q_{num}/q_{ind}$ & $q_{num}/q_{ind}$\\
\hline
10 & 0.998770 & 0.998976\\
20 & 0.999404 & 0.999501\\
50 & 0.999767 & 0.999804\\
100 & 0.999884 & 0.999903\\
200 & 0.999942 & 0.999951\\
500 & 0.999977 & 0.999981\\
1000 & 0.999988 & 0.999990\\
2000 & 0.999994 & 0.999995\\
5000 & 0.999997 & 0.999998\\
\hline

\end{tabular}
\caption{\label{tab:numtest} Tests of the precision of our numerical simulations of the CCE ($q_{num}$) compared with exact analytical results ($q_{ind}$) for varying values of the trapping product, $[n \sigma]^{-1}$. Results are presented for both the anode signal (electron-collecting electrode) and the cathode signal (hole-collecting electrode). The precision of the simulations decreases with increased trapping as expected; however, the precision is more than adequate to simulate the effects on spectroscopy in our germanium detectors (see text for details).}
\end{table}

Before diving into the simulated CCE curves for our GeDs we need to validate the precision of our numerical simulations. The detailed response of the detectors in the presence of trapping for both charge carriers, varying electric fields and consequently drift velocities, and complex weighting fields (Shockley-Ramo theorem) for the collection electrodes must be solved numerically. We have previously developed a custom IDL code to model the charge transport processes and induced signals within our detectors, including detailed modeling of the electric and weighting fields and the electric-field dependencies of the charge carrier drift velocities \cite{amrose2001numerical}. This code simulates the movement of charge carriers over finite time steps ($\Delta t = 0.5$\,ns) along the electric field and utilizes the weighting field to compute the induced signal on the detector electrodes at each step. Previously, this code has been successfully utilized to benchmark the depth calibrations \cite{bandstra2006position} and the cross-talk effects between neighboring electrodes in the COSI-APRA detectors \cite{liu2009characterizing}.  

Building off this modeling effort, we have now incorporated trapping into the charge transport simulations to study the effects of charge trapping (intrinsic and radiation-damage induced) on the CCE, and resulting spectral performance. Separate charge transport simulations that we have previously developed utilized a trapping mean free drift length, $\lambda$, to characterize the effects of radiation damage in the INTEGRAL/SPI co-axial germanium detectors \cite{ho1998pulse}, or utilized a mean trapping time, $\tau$, to characterize the charge transport in CdZnTe detectors similar to those ultimately utilized in the NuSTAR mission \cite{chen2002numerical}. Here we have incorporated the full trapping model into our current simulations, including the effects of both the drift velocities and thermal velocities in calculating the path lengths traveled by the charge carriers. 

In order to confirm the accuracy and precision of our simulations we tested the numerical trapping algorithms on an ideal detector configuration where (1) the contacts were arranged as a simple planar detector of thickness $d = 1.5$\,cm, (2) the electric fields were constant throughout the volume ($400$\,V/cm), and (3) charge trapping was only enabled for one charge carrier at a time.  In this idealized configuration the weighting field is simple to calculate, and both the drift velocities and the thermal velocities of the charge carriers are constant. It is straightforward to calculate the signal induced by the trapped charge carrier on its collection electrode ($q_{ind}$) after the charge cloud has drifted across the entire thickness, $d$, of the detector experiencing charge trapping along the way (e.g., \cite{mei2020impact}). This is a simple variation of the Hecht function \cite{hecht1932mechanismus}. This induced signal is given by the formula:

\begin{equation}
q_{ind} =  \frac{q_0}{l [n \sigma]} (1-e^{-l [n \sigma]});  l = d \sqrt{1+\frac{v_{th}^2}{v_d^2}}
\label{eqn:perfecttrap}
\end{equation}

\noindent While this formula cannot be used directly to model the complex electric and weighting fields of our actual cross-strip detectors, it does enable us to directly test the precision of our numerical charge transport algorithms against an analytical detector response. In Table~\ref{tab:numtest} we present the results of the numerical charge collection signal ($q_{num}$) compared with the analytical expectation ($q_{ind}$, Eqn.~\ref{eqn:perfecttrap}). Given that the spectral resolution for our germanium detectors can reach $E/\Delta E \sim 500$, we would ideally reach an order of magnitude better precision in these numerical simulations to accurately simulate the effects on spectroscopy, i.e., less than 2 parts per 10,000 (i.e., $q_{num}/q_{ind} > 0.9998$). This precision holds for our simulations as long as $[n \sigma]^{-1}$ exceeds $\sim 50$\,cm. However, typical values for germanium appear in the range $[n \sigma]^{-1} \geq 500$\,cm where the simulations achieve a precision of 2 parts per 100,000 (i.e., $q_{num}/q_{ind} > 0.99998$). Therefore, the numerical precision of our charge transport simulation is two orders of magnitude more precise than the anticipated effects on the spectral resolution, and hence more than adequate to model the effects of trapping on our simulated spectra. 

\begin{figure}
\centering
\includegraphics[width=0.6\textwidth]{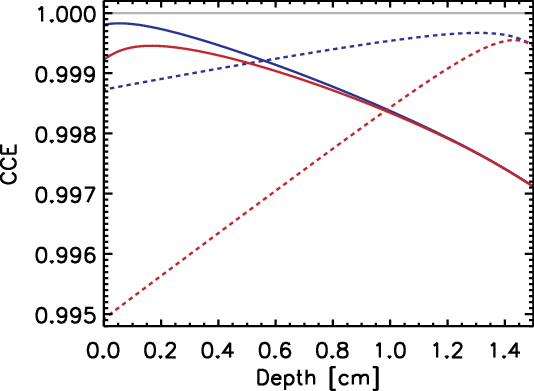}
\caption{\label{fig:f2} Example CCE curves as a function of interaction depth in our detectors. We show both the anode signals (electron-collecting electrode, solid lines), and the cathode signals (hole-collecting electrode, dotted lines). In the first case we simulated $[n \sigma]^{-1}_{e} = 1000$\,cm, $[n \sigma]^{-1}_{h} = 2000$\,cm, representing intrinsic trapping in an undamaged detector (blue lines). In the second case we simulated $[n \sigma]^{-1}_{e} = 1000$\,cm, $[n \sigma]^{-1}_{h} = 500$\,cm, representing a radiation-damaged detector (red lines).}
\end{figure}

\section{Charge Collection Efficiency Curves}
\label{sect:CCE}

In order to demonstrate our numerical simulations we have modeled the charge collection efficiency (CCE) as a function of interaction depth within the detector for one of our specific COSI-APRA detectors (Detector 4). This detector is made of p-type germanium with an impurity concentration of $5.0\times10^{9}$\,cm$^{-3}$, a thickness of $1.5$\,cm, and $2.0$\,mm strip pitch. The detector was operated at a bias of $1000$\,V, exceeding the depletion voltage of $600$\,V.

In Fig.~\ref{fig:f2} we plot our calculated CCE as a function initial interaction depth in the detector, for both the anode signal (solid lines) and the cathode signal (dotted lines). It is worth keeping in mind that due to the small pixel effect the anode signal (i.e., the electron-collecting electrode signal) is dominated by the charge induced by the free electrons drifting towards the anode, but there is still a significant contributions from free holes drifting away towards the cathode. Hence the CCE curve for the anode signal is impacted by both electron trapping and hole trapping. The same is true for the cathode signal (i.e., the hole-collecting electrode signal). Here we chose two cases for demonstration. For the first case we chose $[n \sigma]^{-1}_{e} = 1000$\,cm, $[n \sigma]^{-1}_{h} = 2000$\,cm. This corresponds to the intrinsic trapping situation for an undamaged detector, where the electron trapping is dominant, though the hole trapping  is still significant.  For the second case we chose $[n \sigma]^{-1}_{e} = 1000$\,cm, $[n \sigma]^{-1}_{h} = 500$\,cm, which corresponds to the hypothetical case of a significantly radiation damaged detector where hole trapping dominates. 

Some initial conclusions can be seen from these curves. First, given the variation in the CCEs as a function of depth, trapping can clearly have a significant effect on the overall spectral resolution of a detector when events are combined from all the interaction depths into a single spectrum. Second, due to the small pixel effect, electron trapping predominantly affects the anode signal, and hole trapping predominantly affects the cathode signal; however, the secondary effects of electron trapping on the cathode signal and vice versa are still significant. 

\section{Benchmarking CCE Curves Against Calibrations}
\label{sect:calibration}

\begin{figure}
\centering
\includegraphics[width=1.0\textwidth]{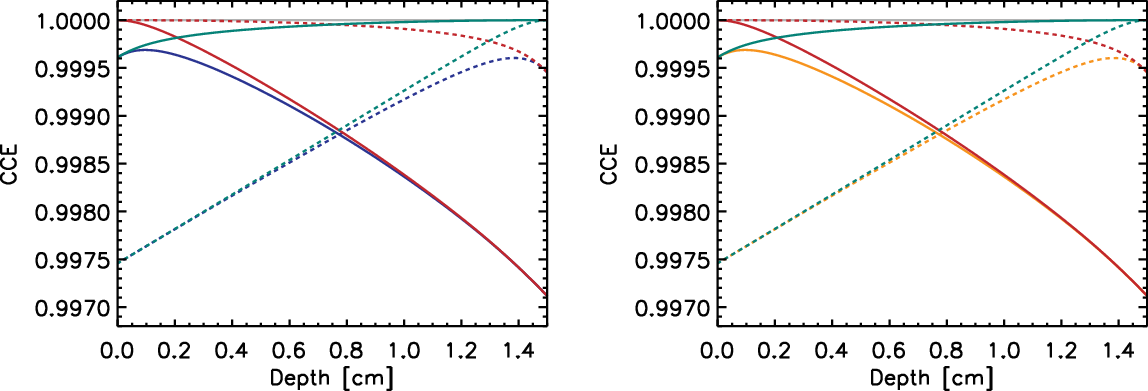}
\caption{\label{fig:f3} (Left) CCE curves for both the anode signals (solid lines), and the cathode signals (dotted lines). We show the resulting CCEs for electron trapping only (no hole trapping) with $[n \sigma]^{-1}_{e} = 1000$\,cm (red curves), hole trapping only (no electron trapping) with $[n \sigma]^{-1}_{h} = 1000$\,cm (teal curves), and trapping for both carriers turned on simultaneously (blue curves).  (Right) The same electron trapping only CCEs (red curves) and hole trapping only CCEs (teal curves) shown to the left, but this plot shows the product of the two CCE curves for each electrode multiplied together (orange curves). These products (orange curves, right panel) are identical to the simulated results where trapping is turned on for both carriers simultaneously (blue curves, left panel).}
\end{figure}

As we have  previously demonstrated \cite{pike2023}, empirical characterization of these CCE curves as a function of depth can be used to correct for the trapping effects and largely restore the spectral resolution at the specific calibration energy used to measure the CCE. A drawback of this previous empirical approach is that it can only provide the relative value of the CCE curves as a function of depth, not the absolute value. This can be seen in Fig.~\ref{fig:f2} where the CCE curves always fall below unity, by an amount that is unknown except through these simulations. Without the absolute value, the empirical corrections can only be utilized over a small energy range centered on the calibration energy and not across the entire energy range of the detector. The numerical simulations presented here can be utilized to better model these CCE curves including the absolute value, as well as predict how the CCE curves will evolve as a detector is radiation damaged. Benchmarking these simulations against calibrations could be a challenging task, perhaps requiring many simulations over multiple values of $[n \sigma]^{-1}_{e}$ and $[n \sigma]^{-1}_{h}$ to find the best fits to measured data. However, we have identified through the simulations two simplifying factors that were not obvious \emph{a priori}, and which help identify a path for calibrating these CCE curves. 

The first simplifying factor is that the effects of electron trapping and hole trapping on the CCE curves are multiplicative. This is demonstrated in Fig.~\ref{fig:f3}. In the left panel we show the CCE curves for the case where $[n \sigma]^{-1}_{e} = 1000$\,cm, $[n \sigma]^{-1}_{h} = 1000$\,cm (blue curves). We also show the curves where only electron trapping is turned on ($[n \sigma]^{-1}_{e} = 1000$\,cm, $[n \sigma]^{-1}_{h} \rightarrow \infty$, red curves), and only hole trapping is turned on ($[n \sigma]^{-1}_{e} \rightarrow \infty$, $[n \sigma]^{-1}_{h} = 1000$\,cm, teal curves). In the right panel we again show the CCE curves where only one charge trapping is turned on at a time (red curves, teal curves), as well as the product of the resulting CCE curves (orange curves). The product of the individual CCE curves (orange curves, right panel) are identical to the CCE curves with trapping turned on for both carriers simultaneously (blue curves, left panel). This gives us the very useful result that the trapping can be modeled and characterized for each of the charge carriers separately, and the combined effect can be calculated by multiplying together the individual CCE curves. 

\begin{figure}
\centering
\includegraphics[width=1.0\textwidth]{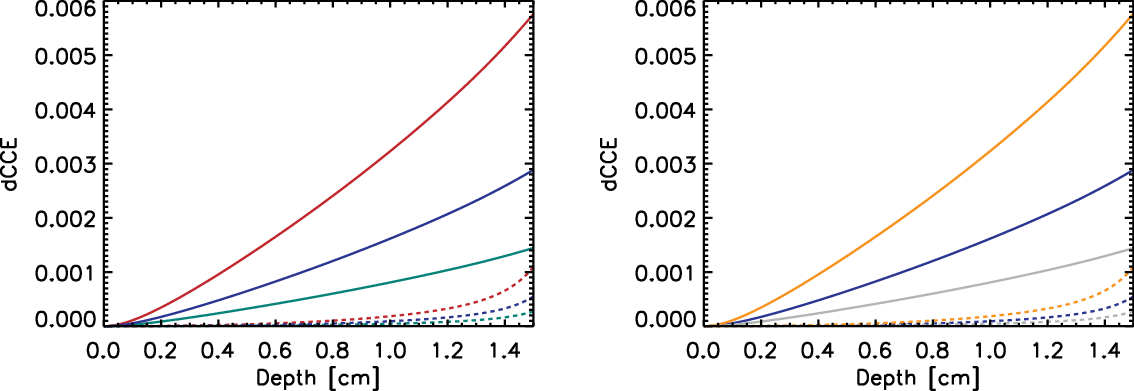}
\caption{\label{fig:f4} (Left) dCCE curves for both the anode signals (solid lines), and the cathode signals (dotted lines) for electron trapping only (no hole trapping) with $[n \sigma]^{-1}_{e} = 500$\,cm (red curves), $1000$\,cm (blue curves), and $2000$\,cm (teal curves). (Right) The same electron trapping only curves with $[n \sigma]^{-1}_{e} = 1000$\,cm (blue curves), but we scale this dCCE by factors of $2$ (orange curves) and $0.5$ (grey curves). These scaled curves (orange and grey curves) are identical to the unscaled curves with corresponding variations in $[n \sigma]^{-1}$ (red and teal curves, left panel).}
\end{figure}

The second simplifying factor is that the trapping effects on the CCE curves scale proportional to $[n \sigma]$ (or inversely proportional to $[n \sigma]^{-1}$). This is demonstrated in Fig.~\ref{fig:f4} where we plot the deviation of the CCE curves from unity, the ``delta CCE'' curves:

\begin{equation}
dCCE(z) \equiv 1 - CCE(z)
\label{eqn:dCCE}
\end{equation}

\begin{figure}
\centering
\includegraphics[width=1.0\textwidth]{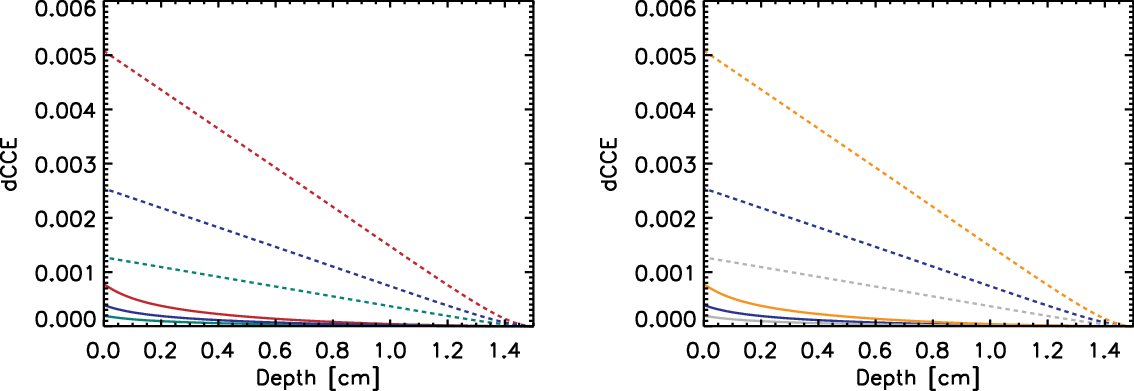}
\caption{\label{fig:f5} Same dCCE curves as in Fig.~\ref{fig:f4}, but for hole trapping only (no electron trapping).}
\end{figure}

\noindent On the left hand side, we plot the numerically calculated curves for the case of electron trapping only, for three values of $[n \sigma]^{-1}_{e} = [500, 1000, 2000$\,cm$]$. The shape of these curves appear identical, just differing by an overall scale factor. On the right hand side we plot the curves for $[n \sigma]^{-1}_{e} = 1000$\,cm, as well as these same curves scaled by a factor of $2$ (corresponding to $500$\,cm) and $0.5$ (corresponding to $2000$\,cm). The resulting curves are identical to those calculated on the left. The same is true for the curves for the case of hole trapping only (Fig.~\ref{fig:f5}). This insight gives us the useful result that the trapping effects can be modeled at a single value of $[n \sigma]^{-1}$ separately for each charge carrier (say $[n \sigma]^{-1} = 1000$\,cm), and then the resulting dCCE curve templates can be scaled directly to different trapping values by multiplying by the ratio of $[n \sigma]$.

In hindsight, these two simplifying assumptions are due to the fact that we are working in the ``weak'' trapping limit for these germanium detectors, as discussed at the end of Section~\ref{sect:model}. In this limit we can ignore second- and higher-order terms in the trapping equations which linearizes the impacts of trapping. This can be seen by reconsideration of the Hecht function, Equation~\ref{eqn:perfecttrap}, in the limit that ${l [n \sigma]} << 1$. In this limit, the induced signal is given by:

 \begin{equation}
q_{ind} =  {q_0}(1-\frac{l [n \sigma]}{2});  {l [n \sigma]} << 1
\label{eqn:limittrap}
\end{equation}

\noindent This limiting case which holds for trapping in germanium detectors explains how the effects of $[n \sigma]^{-1}$ become linearized and result in the two simplifying factors we have identified in this section.

These two simplifying factors provide us with a convenient means of benchmarking measured data against the simulations. The first step is to simulate the CCE curve templates (and the resulting dCCE curves) separately for electron trapping and hole trapping, assuming a fixed value of  $[n \sigma]^{-1}$ (say $1000$\,cm) for each case. The second step is to fit the measured CCE curves to the simulations by scaling the simulated dCCE curves for each carrier trap, subtract the dCCE curves from unity to calculate the resulting CCE curves for each carrier trap, and multiplying the resulting CCEs together to form the overall CCE curves, realizing that $[n \sigma]^{-1}_{e}$ and $[n \sigma]^{-1}_{h}$ are jointly constrained by both the anode signal CCE and the cathode signal CCE. We demonstrate such a fit to measured data in Section~\ref{sect:data}.

\section{Averaging Initial Interactions Across Electrodes}
\label{sect:average}

\begin{figure}
\centering
\includegraphics[width=0.7\textwidth]{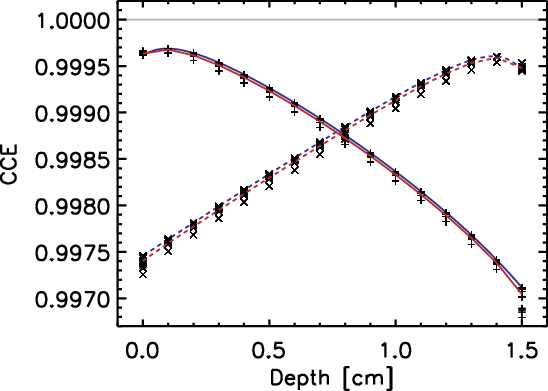}
\caption{\label{fig:f6} CCE curves for both the anode signals (solid lines), and the cathode signals (dotted lines). At each depth, $z$, in the detector we sampled an array of $11 \times 11$ initial interaction sites in $x$ and $y$ to determine the variation of the CCE at a given depth. The scatter of these 121 points at each depth is small enough to be mostly indistinguishable in the plot beyond a few outlier points corresponding to the edges of the electrodes. Shown for comparison are the CCE curves for the electrode center (blue curves) and the CCE curves derived from averaging the values for difference initial interaction locations spanning the widths of the electrodes (red curves). This exercise shows that the CCE curves do not vary significantly across the electrodes.}
\end{figure}

So far in this paper we have only shown CCE curves derived for interactions \emph{centered} on the signal cathode and anode. We anticipate that the CCE curves may differ when the variation of the initial interaction locations across the electrodes (for a given depth) are taken into account. To study the effect of varying the initial interaction location at a given depth we sampled an $11 \times 11$ array of points in $x$ and $y$, spanning the width of the signal strips. In Fig.~\ref{fig:f6} we show the scatter of CCE values at each depth within the detector, as well as the resulting CCE curve found by averaging these values at each depth. 

The scatter plot shows little variation in CCE compared with the CCE curves derived from interactions in the center of the electrodes. The bulk of the scatter points (111 points at each depth) fall squarely on the initial CCE curves (blue lines), with a handful of outliers falling at slightly lower CCE values. When we average over all of these points for a given depth, the resulting average CCE curves (red lines) fall only slightly below the original CCE curves (blue lines), and would be indistinguishable for all but the most detailed calibrations. From this study we conclude that utilizing the CCE curves derived for interactions centered on the signal electrodes are adequate for benchmarking simulations against calibration data. 

\section{Comparison with Measurements}
\label{sect:data}

\begin{figure}
\centering
\includegraphics[width=0.7\textwidth]{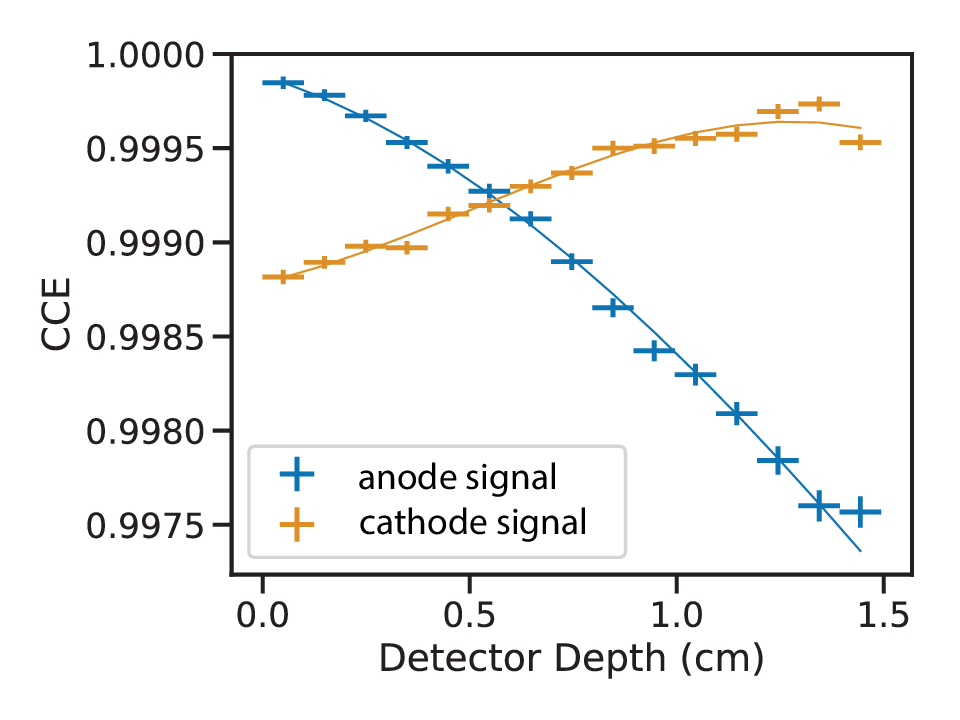}
\caption{\label{fig:f7} Measured CCE curves fitted to simulations. We find that the simulations match the measurements well and allow for the extraction of the electron and hole trapping products specific to the detector.}
\end{figure}

We performed comparison of our simulated CCE curves against data collected during the 2016 calibration campaign of the GeDs aboard the COSI balloon payload. The details of these calibrations can be found in \cite{beechert2022calibrations}. Here we show the measured CCE curves for one COSI-APRA GeD corresponding to Detector\,4 in \cite{kierans2018detection}, the same detector we modeled in Figure \ref{fig:f2}. In particular, we investigated the $662$\,keV emission line collected using a $^{137}$Cs source, considering only events which triggered a single pixel (i.e., an individual signal electrode on each face of the detector). Using the results of the calibration campaign, we applied a first-order energy calibration in order to map the measured pulse height amplitude to photon energy for each interaction event. This gain correction did not take into consideration the depth of photon interaction, meaning that information about the CCE as a function of depth is preserved.

We considered all single-pixel events across the detector, sorting events into 15 bins according to depth of interaction. Each event is tagged with two values for the inferred photon energy using the first-order energy calibration, corresponding to the values measured using the electron and hole signals. We thus constructed two lists of event energies, and for each list we excluded events with inferred photon energy outside the range $640\,\mathrm{keV}<E<672\,\mathrm{keV}$ in order to include only photopeak events. For each $1$\,mm depth bin, we fitted the unbinned event energies to the spectral line profile model described in \cite{boggs2023b}, which provides an unbiased measurement of the inferred centroid energy of the emission line by accurately modeling the low-energy tailing which results from charge sharing across adjacent pixels. We fitted the data to the line profile models by constructing a cost function for unbinned Poisson-distributed data \cite{pike2023} and minimizing this function using the Python package iminuit \cite{iminuit}, which utilizes the minimization algorithm described in \cite{james1975minuit}. We thereby determined the inferred centroid energy of the line profile measured at each depth bin. This peak fitting procedure is identical to the one which we describe in more detail in \cite{pike2023}.

Next, we fitted the two measured lists of depth-dependent centroid energies to the simulated CCE curves. We utilized the simplifying factors described in Section \ref{sect:calibration} in order to write a model for the centroid energy, $E$, as a function of depth, $z$:

\begin{equation}\label{eqn:CCE_energy}
    E_{e,h}(z) = A_{e,h} \times \left[1-B\times dCCE_{e,h}|_e(z)\right]\times\left[1-C\times dCCE_{e,h}|_h(z)\right]
\end{equation}

\noindent where the subscripts $e$ and $h$ indicate the anode (electron-collecting electrode) and cathode (hole-collecting electrode) signals, respectively. We introduce normalization factors, $A_{e,h}$, in order to account for the first-order energy calibration which independently maps the anode signal and cathode signal CCE curves into energy space (thus, the CCE is given by $E_{e,h}(z)/A_{e,h}$). The simulated dCCE curves assuming only electron trapping are represented by $dCCE_{e,h}|_e(z)$, while the simulated dCCE curves assuming only hole trapping are represented by $dCCE_{e,h}|_h(z)$. For all CCE curve templates, the trapping product was set to a value of $[n\sigma]^{-1}=1000$\,cm. The factors $B$ and $C$ represent the scaling factors of the electron and hole trapping, respectively. 

We performed simultaneous least-squares fitting of $E_e(z)$ to the list of centroid energies determined using the anode signal and $E_h(z)$ to the list of centroid energies determined using the cathode signal. In this way, we fit for a total of four parameters: $A_e$, $A_h$, $B$, and $C$. The results of this least-squares fitting are shown in Figure \ref{fig:f7}. We have divided the measured centroid energies by their corresponding normalizations $A_{e,h}$ in order to show the underlying CCE curves. We find that the measured data fit the simulations well, achieving a reduced Chi-squared of $\chi^2_{\nu}=1.3$. Our fits also allow us to extract the trapping products for each species of charge carrier. For this detector, we find $B=0.98\pm0.02$ and $C=0.49\pm0.01$, corresponding to an electron trapping product of $[n\sigma]^{-1}_e=1020\pm20$\,cm and a hole trapping product of $[n\sigma]^{-1}_h=2040\pm50$\,cm.

Finally, we utilized the best-fit CCE curves to correct for the trapping effects on the spectra, which us identical to the empirical method employed in \cite{pike2023}, except utilizing the numerical CCE curves instead of empirical CCE curves. The resulting improvements in the spectral performance are identical to those presented in  \cite{pike2023}.

\section{Discussion}
\label {sect:disc}

We anticipate over the course of the COSI space mission that the exposure to damaging radiation (primarily energetic protons) will induce increasing levels of charge trapping (primarily hole trapping) in the germanium strip detectors. Even though radiation damage has the potential to significantly increase the charge trapping, the GeDs on the COSI mission should always remain in the limit where $[n \sigma]^{-1} >> d$ (Section~\ref{sect:model}), enabling us to continuously correct for the trapping over time to largely restore and maintain the spectral resolution of the detectors. The ability to characterize the CCE curves as a function of interaction depth in these GeDs enables the powerful capability of correcting for the effects of charge trapping to maintain the optimal spectral resolution of the detectors, even as the trapping effects increase over time. 

In this paper we have demonstrated techniques for modeling the CCE curves for a baseline trapping product ($[n\sigma]^{-1}=1000$\,cm). We utilized the resulting CCE curve templates to benchmark the simulations against measured data, characterizing the underlying intrinsic trapping products for the detector. The most powerful aspect of this work is that the underlying CCE curve templates themselves do not change for a given detector, even as $[n\sigma]^{-1}$ evolves over time due to increased damage. As long as we can characterize $[n\sigma]^{-1}$ as a function of time over the mission, either empirically or through ancillary measurements, we will be able to implement a time-dependent spectral correction for each detector.

The simulations and benchmarking techniques presented in this work will be utilized in several ways in the COSI program. First, the simulated CCE curve templates will enable us to measure and correct the intrinsic trapping in both the existing COSI-APRA detectors as well as the COSI-SMEX detectors as they come on line, optimizing the spectral performance of these detectors. Second, we have a program to radiation damage COSI-APRA detectors in a proton beamline which will enable us to characterize the induced trapping as a function of proton fluence by benchmarking the resulting CCE curves using the methods developed here. Finally, the techniques presented here will enable us to characterize and correct for radiation damage effects for COSI-SMEX as they evolve over the lifetime of the mission, helping to optimize and maintain the spectral resolution of the instrument. 

\section{Acknowledgements}
\label{}

This work was supported by the NASA Astrophysics Research and Analysis (APRA) program, grant 80NSSC22K1881. Thank you to the manuscript reviewers for their helpful suggestions.

\bibliographystyle{elsarticle-num-names} 
\bibliography{refs}

\end{document}